\documentclass[10pt,a4paper]{article}

\usepackage[latin1]{inputenc} 
\usepackage[dvips]{graphicx} 
\usepackage{a4wide}
\usepackage{latexsym}
\usepackage{amsmath}
\usepackage{amssymb}

\title{$\Theta^+$ width estimation with nonzero momentum transfer}

\author{Cédric Lorcé\\ \small{\emph{Université de
Liège, Institut de Physique, Bât. B5a, B4000 Liège, Belgium}}\\
\small{\emph{Ruhr-Universität Bochum, Institut für Theoretische Physik II, D-44780 Bochum, Germany}}\\
\small{\emph{E-mail: C.Lorce@ulg.ac.be}}}
\date{}

\newcommand{\ud}{\mathrm{d}}

\newcommand{\uM}{\mathcal{M}}
\newcommand{\uPcal}{\mathcal{P}}
\newcommand{\uZ}{\mathcal{Z}}
\newcommand{\uQcal}{\mathcal{Q}}

\newcommand{\up}{\mathbf{p}}
\newcommand{\uP}{\mathbf{P}}
\newcommand{\uq}{\mathbf{q}}

\newcommand{\uD}{\mathcal{D}}
\newcommand{\uV}{\mathbf{V}}
\newcommand{\uVcal}{\mathcal{V}}

\begin{document}

\maketitle

\begin{center}
\begin{minipage}[t]{15cm}
\small{We have used the light-cone formulation of Chiral-Quark
Soliton Model to estimate the width of the lightest pentaquark
$\Theta^+$. We have found that the effect of nonzero momentum
transfer is important and reduces drastically the width to about
$0.43$ MeV. This means that this effect is a piece of the small
width puzzle of exotic baryons.}
\end{minipage}
\end{center}

\section{Introduction}

Chiral-Quark Soliton Model ($\chi$QSM) has been recently formulated
in the infinite momentum frame (IMF) \cite{PetPol,DiaPet}. This
provides a new approach for extracting predictions out of the model.
The light-cone formulation is attractive in many ways. For example
light-cone wave functions are particularly well suited to compute
matrix elements of operators. One can even choose to work in a
specific frame where the annoying part of currents, \emph{i.e.} pair
creation and annihilation part, does not contribute. On the top of
that it is in principle also easy to compute parton distributions
once light-cone wave functions are known.

The technique has already been used to study vector and axial
charges of the nucleon and $\Theta^+$ pentaquark width up to the
5-quark component \cite{DiaPet,Moi}. It has been shown that
relativistic effects are non-negligible. For example they explain
the reduction of the naïve quark model value $\frac{5}{3}$ for the
nucleon axial charge $g^{(3)}_A$ down to a value close to $1.257$
observed in beta decays.

The existence of an exotic antidecuplet is still under debate. Even
though most of the latest experiments suggest that it does not
exist, no definitive answer can be given \cite{Azimov}.
Theoretically pentaquarks are expected to have narrow width. In
their seminal paper \cite{DPP}, Diakonov, Petrov, Polyakov have
given an upper bound $\Gamma_{\Theta^+}\leq 15$ MeV. In a previous
paper using the present technique \cite{Moi} we have obtained
$\Gamma_{\Theta^+}\sim 2$ MeV. On the experimental side, if
$\Theta^+$ does exist, its width should be
$\Gamma_{\Theta^+}=0.36\pm 0.11$ MeV \cite{ExpWidth}. Such a small
value is below experimental resolution and does not contradict any
other experimental result on $\Gamma_{\Theta^+}$. More conservative
phenomenological estimations give only upper bounds of 1-5 MeV
\cite{Pheno}. Using the so-called ``model-independent approach'' to
$\chi$QSM \cite{Ghil} it has been shown that the model may be
consistent with $\Gamma_{\Theta^+}<1$ MeV and even the experimental
value. Our estimation \cite{Moi} is one order of magnitude higher
than the DIANA result. This can be related to the fact that we did
not take into account the difference of masses between nucleon and
$\Theta^+$. The axial matrix element was then evaluated at zero
momentum transfer. That is the reason that motivated this study.

In this paper we present our results for nonzero momentum transfer.
We show that the effect is far from being negligible and thus is
probably part of the explanation for the small width of pentaquarks.
While in a previous work we have considered relativistic corrections
to quark wavefunction, in this study we limit ourselves to the
nonrelativistic case so that the computations remain tractable in a
reasonable amount of time. In section \ref{Section deux} we show how
we have considered nonzero momentum transfer in this study. In
section \ref{Section trois} we explain how to compute the matrix
elements within the $\chi$QSM in IMF. Then we give the results
obtained and the numerical values of relevant integrals in section
\ref{Section quatre}.

\section{Nonzero momentum transfer}\label{Section deux}

Were nucleon and $\Theta^+$ degenerate in mass there would be no
momentum transfer\footnote{We work in the Drell frame $q^+=0$ where
quark-antiquark creation and annihilation are absent. The momentum
transfer can then only be transverse.}. Let us consider a $\Theta^+$
pentaquark with 4-momentum $\uPcal$ decaying into a nucleon and a
kaon with 4-momenta $\uPcal'$ and $q$ respectively. We consider the
$z$-direction as the pentaquark momentum one and all the particles
on mass-shell
\begin{eqnarray}
\uPcal&=&\left(\sqrt{\uPcal_z^2+M^2},\vec 0,\uPcal_z\right),\\
\uPcal'&=&\left(\sqrt{X^2\uPcal_z^2+q_\perp^2+M'^2},-\vec q_\perp, X \uPcal_z\right),\\
q&=&\left(\sqrt{(1-X)^2\uPcal_z^2+q_\perp^2+m^2},\vec q_\perp, (1-X)
\uPcal_z\right)
\end{eqnarray}
where $M$, $M'$ and $m$ are respectively the $\Theta^+$, nucleon and
kaon masses and $X$ the fraction of the total longitudinal momentum
kept by the nucleon. In the Infinite Momentum Frame (IMF)
$\uPcal_z\to\infty$ the energy conservation law yields the following
condition
\begin{equation}\label{Constraint}
M^2=\frac{M'^2+q_\perp^2}{X}+\frac{m^2+q_\perp^2}{1-X}.
\end{equation}

The nucleon is described as a superposition of $3+2n$-quark Fock
components with $n=0,1,2,\ldots$ while a pentaquark has
$n=1,2,\ldots$. The momenta of the individual quarks have to sum up
to the total momentum of the baryon they belong to
\begin{equation}
\sum_{i=1}^{3+2n}\vec{p}_i=\vec{\uPcal}.
\end{equation}
We introduce $z_i=p_{iz}/\uPcal_z$ the fraction of the total
longitudinal momentum carried by quark $i$. The two other components
of the momentum are collectively called $\vec{p}_{i\perp}$. This
means that
\begin{equation}\label{Conditions}
\sum_{i=1}^{3+2n}\vec{p}_{i\perp}=\vec{\uPcal}_\perp\qquad\textrm{and}\qquad
\sum_{i=1}^{3+2n}z_i=1.
\end{equation}
Using eq. (\ref{Conditions}) one concludes that
\begin{equation}
\sum_{i=1}^{3+2n}\vec{p}_{i\perp}=\vec{0},\quad
\sum_{i=1}^{3+2n}z_i=1\quad\textrm{and}\quad
\sum_{i=1}^{3+2n}\vec{p}'_{i\perp}=-\vec{q}_\perp,\quad
\sum_{i=1}^{3+2n}z'_i=1
\end{equation}
where the unprimed variables refer to pentaquark and primed ones to
nucleon. The current strikes only one quark line, say $j_0$, so one
obtains the following relations
\begin{eqnarray}\label{Connection}
\vec{p}_{j\perp}=\vec{p}'_{j\perp},&\qquad&\vec{p}_{j_0\perp}=\vec{p}'_{j_0\perp}+\vec{q}_\perp,\\
z_j=X z'_j,&\qquad&z_{_0}=X z'_{j_0}+(1-X)
\end{eqnarray}
where $j\neq j_0$ refers to the non-struck quarks. Since in the IMF
all the quarks are moving in the same direction as the baryon they
belong to, one expects $z_i,z'_i\in[0,1]$. In fact, from eq.
(\ref{Connection}) one can see that in order for the pentaquark to
decay any internal configuration is not allowed and depends on the
fraction of total momentum carried by the nucleon
\begin{equation}
z_j\in[0,X]\qquad\textrm{and}\qquad z_{j_0}\in[1-X,1].
\end{equation}
Any value of $X$ is also not allowed because of the energy
constraint (\ref{Constraint}). In this study we have considered the
cases of a massless $m=0$ and a massive $m=495$ MeV kaon and have
used $M=1530$ and $M'=938$ MeV yielding to $X\in[0.376,1]$ and to
$X\in[0.468,0.803]$ respectively. In the equal mass limit $M=M'$
with massles kaon $m=0$ one has $X=1$ and $\vec{q}_\perp=0$ as it
should be.

\section{Nonzero momentum transfer integrals}\label{Section trois}

It was shown in \cite{DiaPet} that four integrals were needed to
compute the axial charge of the $\Theta\to K N$ decay. If one
consider nonzero (transverse) momentum transfer then one has to
compute five integrals\footnote{The nonzero momentum transfer breaks
the symmetry of the quark-antiquark pair leading to $K_{3\sigma}\neq
K_{\sigma 3}$.}. They can be written in the general form
\begin{equation}
K_I=\frac{M^2}{2\pi}\int \ud
X\frac{\ud^3\uP'}{(2\pi)^3}\,\Phi_I(\uP',X)\,G_I(\uP',X)
\end{equation}
where $\Phi_I$ is a valence probability distribution, $G_I$ is a
quark-antiquark probability distribution and
$I=\pi\pi,\sigma\sigma,33,\sigma 3,3\sigma$. These integrals are
regularized by means of Pauli-Villars procedure. In order to keep
them tractable in a reasonable amount of time, we used the
nonrelativistic form of the valence probability distribution
\begin{eqnarray}
\Phi_I(\uP',X)&=&\int\ud
z'_{1,2,3}\frac{\ud^2\up'_{1,2,3\perp}}{(2\pi)^6}\,\delta(\frac{P'_z}{\uM}+z'_1+z'_2+z'_3-1)(2\pi)^2\delta^{(2)}(\uP'_\perp+\up'_{1\perp}+\up'_{2\perp}+\up'_{3\perp}+\uq_\perp)\nonumber\\
&\times&h(p_1)h(p'_1)h(p_2)h(p'_2)h(p_3)h(p'_3)
\end{eqnarray}
with $P'_z=(z'_4+z'_5)\uM=Z'\uM$ and
$\uP'_\perp=\up'_{4\perp}+\up'_{5\perp}$. More details about the
expressions used can be found in \cite{DiaPet,Moi}.

\subsection{Struck valence quark integrals $I=\pi\pi,\sigma\sigma,33$}

If the struck quark is a valence one then one can choose $j_0=3$ in
(\ref{Connection}). The quark-antiquark probability distributions
are
\begin{eqnarray}
G_{\pi\pi}(\uP',X)&=&\theta(P'_z)P'_z\Pi(\uP)\Pi(\uP')\int^1_0\ud
y\int\frac{\ud^2\uQcal_\perp}{(2\pi)^2}\frac{\uQcal^2_\perp+M^2}{\uZ\uZ'},\label{Integrals
begin}
\\
G_{\sigma\sigma}(\uP',X)&=&\theta(P'_z)P'_z\Sigma(\uP)\Sigma(\uP')\int^1_0\ud
y\int\frac{\ud^2\uQcal_\perp}{(2\pi)^2}\frac{\uQcal^2_\perp+M^2(2y-1)^2}{\uZ\uZ'},
\\
G_{33}(\uP',X)&=&\theta(P'_z)\frac{P_z{P'_z}^2}{\uP\uP'}\Pi(\uP)\Pi(\uP')\int^1_0\ud
y\int\frac{\ud^2\uQcal_\perp}{(2\pi)^2}\frac{\uQcal^2_\perp+M^2}{\uZ\uZ'}
\end{eqnarray}
where $\uZ=\uQcal^2_\perp+M^2+y(1-y)\uP^2$ and
$\uZ'=\uQcal^2_\perp+M^2+y(1-y)\uP'^2$. The internal quark-antiquark
pair variables are defined as $y=z'_5/Z'$ and
$\uQcal_\perp=y\up'_{5\perp}-(1-y)\up'_{4\perp}$. One naturally
recovers the zero momentum transfer case by setting $X=1$ and thus
$\uq_\perp=\mathbf{0}$.

\subsection{Struck quark/antiquark integrals $I=\sigma 3,3\sigma$}

If the struck quark is the quark (antiquark) of the pair one has
$j_0=4$ ($j_0=5$) in (\ref{Connection}). Let us consider that the
current strikes the antiquark. The quark-antiquark probability
distributions are then
\begin{eqnarray}
G_{\sigma
3}(\uP',X)&=&\theta(P'_z)\frac{{P'_z}^2}{\uP'}\Sigma(\uP)\Pi(\uP')\int^1_0\ud
y\int\frac{\ud^2\uQcal_\perp}{(2\pi)^2}\frac{(\uQcal^2_\perp-M^2)V_z+(1-y)(\uQcal_\perp\cdot\uV_\perp+2M^2
Z')}{\uD\uZ'},
\\
G_{3\sigma}(\uP',X)&=&\theta(P'_z)\frac{P_zP'_z}{\uP}\Pi(\uP)\Sigma(\uP')\int^1_0\ud
y\int\frac{\ud^2\uQcal_\perp}{(2\pi)^2}\frac{(\uQcal^2_\perp-M^2(2y-1))V_z+(1-y)\uQcal_\perp\cdot\uV_\perp}{\uD\uZ'}
\label{Integrals end}
\end{eqnarray}
where
\begin{equation}
\uD=(1-y)\frac{V_z\uVcal_z}{Z'}(\uM^2X^2Z'^2+\uP^2_\perp)+V_z(\uQcal^2_\perp+M^2)+(1-y)\frac{1}{Z'}\uV^2_\perp+\frac{2}{Z'}(1-y)(\uVcal_z\uP_\perp+Z'\uQcal_\perp)\cdot\uV_\perp.
\end{equation}
For convenience we have introduced the variables
$V_z=Z'+\frac{1-X}{X}$, $\uVcal_z=y Z'+\frac{1-X}{X}$ and
$\uV_\perp=Z'\uq_\perp-\frac{1-X}{X}\uP_\perp$. Once more one
naturally recovers the zero momentum transfer case
$K_{3\sigma}=K_{\sigma 3}$ by setting $X=1$ and thus
$\uq_\perp=\mathbf{0}$.

If the current stroke the quark we would in fact obtain the same
quark-antiquark probability distributions since $\uP'$ and $X$ do
not change under a permutation of indices 4 and 5. If we had started
with the quark struck we would have ended up with the same
expression after a change of variables $y\to(1-y)$ and
$\uQcal_\perp\to-\uQcal_\perp$ corresponding indeed to a permutation
of indices 4 and 5.

\section{Numerical results}\label{Section quatre}

In the evaluation of the scalar integrals we have used the quark
mass $M=345$ MeV, the Pauli-Villars mass $M_\textrm{PV}=556.8$ MeV
for the regularization and the baryon mass $\uM=1207$ MeV as it
follows for the ``classical'' mass in the mean field approximation
\cite{Approximation}.

The numerical evaluation for $m=0$ yields
\begin{equation}\label{Integral values}
K_{\pi\pi}=0.02198,\qquad K_{\sigma\sigma}=0.00883,\qquad
K_{33}=0.01216,\qquad K_{3\sigma}=0.01105,\qquad K_{\sigma
3}=0.01066.
\end{equation}

The numerical evaluation for $m=495$ MeV yields
\begin{equation}\label{Integral values}
K_{\pi\pi}=0.01645,\qquad K_{\sigma\sigma}=0.00630,\qquad
K_{33}=0.00907,\qquad K_{3\sigma}=0.00799,\qquad K_{\sigma
3}=0.00683.
\end{equation}

From these integrals one can estimate the width of $\Theta^+$ width
as explained in \cite{DiaPet}. In table \ref{Result} we compare the
results for $\Theta^+$ width with zero and nonzero momentum
transfer.
\begin{table}[h!]\begin{center}\caption{\small{Estimation of $\Theta^+$ width with and without momentum transfer.\newline}}
\begin{tabular}{c|c|cc}
\hline\hline \rule{0pt}{2.5ex}&\multicolumn{1}{c|}{$\uq_\perp=0$}&\multicolumn{2}{c}{$\uq_\perp\neq 0$}\\
&$M=M'$, $m=0$&$M\neq M'$, $m=0$&$M\neq M'$, $m\neq 0$\\
\hline\rule{0pt}{3ex}
$g_A(\Theta\to KN)$&0.202&0.063&0.042\\
$g_{\Theta KN}$&2.230&0.697&0.467\\
$\Gamma_\Theta$ (MeV)&4.427&0.432&0.194\\ \hline\hline
\end{tabular}\label{Result}\end{center}
\end{table}
Not surprisingly one can see that nonzero momentum transfer reduces
the width. Indeed one can expect the integrals to be smaller because
all the configurations for the pentaquark to decay are not allowed.
Of course one should not trust the number of Table \ref{Result} as
they are because up to now we did not estimate the theoretical
errors. Nevertheless we hope that they give some kind of order of
magnitude. Note that it is in rather good agreement with the
experimental extraction.

\section{Conclusion}

The question of the pentaquark is a very interesting one since its
existence or non-existence would should light on many aspects and
problems of baryon physics and low-energy QCD. One of the most
interesting question is its width. While there is no definitive
theoretical answer, experiments seem to suggest a very small width
($<$1 Mev!) if it does exist. It is thus imperative to see if one
can obtain such a small value within a model and try to understand
the reason.

Using $\chi$QSM formulated in the Infinite Momentum Frame (IMF) we
were able to give an estimation $\approx 2$ MeV. Such a small values
was attributed to the fact in IMF the current does not create nor
annihilate quark-antiquark pairs. So the pentaquark can only be
connected the 5-quark Fock component of the nucleon, which is small
compared the the 3-quark Fock component.

In this paper we tried to take into account the fact that $\Theta^+$
and $N$ have different masses and thus that the current has nonzero
(transverse) momentum. We have obtained $\Gamma_{\Theta^+}\approx
0.43$ MeV which is of the same order of magnitude than the
experimental width $\Gamma_{\Theta^+}=0.36\pm 0.11$ MeV. We have
shown that this can be understood by the fact that a nonzero
momentum transfer reduces the number of possible configurations for
the pentaquark to decay. Although our value cannot be fully trusted
(due to unknown theoretical errors) it is an indication that nonzero
momentum transfer is partly responsible for such a small width.

\subsection*{Acknowledgements}

The author is grateful to RUB TP2 for its kind hospitality, to D.
Diakonov for suggesting this work and to M. Polyakov for his careful
reading and comments. The author is also indebted to J. Cugnon whose
absence would not have permitted the present work to be done. This
work has been supported by the National Funds of Scientific
Research, Belgium.

\end{document}